\documentclass{article}

\usepackage{arxiv}

\usepackage[utf8]{inputenc} 
\usepackage[T1]{fontenc}    
\usepackage{hyperref}       
\usepackage{url}            
\usepackage{booktabs}       
\usepackage{amsfonts}       
\usepackage{nicefrac}       
\usepackage[ruled,vlined]{algorithm2e}
\usepackage{microtype}      
\usepackage{graphicx}
\title{Circuitscape in Julia: High Performance Connectivity Modelling to Support Conservation Decisions}

\author{
    Ranjan Anantharaman\\
    Massachusetts Institute of Technology\\
    Cambridge, MA \\
    \texttt{ranjanan@mit.edu}
    \AND
    Kimberly Hall\\
    The Nature Conservancy \\
    Lansing, MI\\
    \texttt{kimberly.hall@tnc.org}
    \AND
    Viral Shah \\
    Julia Computing Inc.\\
    Cambridge, MA \\
    \texttt{viral@juliacomputing.com}
    \AND
    Alan Edelman\\
    Massachusetts Institute of Technology\\
    Cambridge, MA \\
    \texttt{edelman@math.mit.edu}
}

\begin{document}
\maketitle

\begin{abstract}
\begin{enumerate}
    \item Connectivity across landscapes influences a wide range of conservation-relevant ecological processes, including species movements, gene flow, and the spread of wildfire, pests, and diseases. Recent improvements in remote sensing data suggest great potential to advance connectivity models, but computational constraints hinder these advances. 

    \item To address this challenge, we upgraded the widely-used Circuitscape connectivity package to the high performance Julia programming language. Circuitscape.jl allows users to solve problems faster via improved parallel processing and solvers, and supports applications to larger problems (e.g., datasets with hundreds of millions of cells). 

    \item We document speed improvements of up to 1800\%. We also demonstrate scaling of problem sizes up to 437 million grid cells.  These improvements allow modelers to work with higher resolution data, larger landscapes and perform sensitivity analysis effortlessly. These improvements accelerate the pace of innovation, helping modelers address pressing challenges like species range shifts under climate change.  
    
    \item Our collaboration between ecologists and computer scientists has led to the use of connectivity models to inform conservation decisions. Further, these next generation connectivity models will produce results faster, facilitating stronger engagement with decision-makers.

\end{enumerate}
\end{abstract}


\section{Introduction}
Connectivity models provide important insights into ecological processes that involve variation in movement or flow patterns across heterogeneous environments \cite{crooks2006}. In applied conservation, connectivity maps are incorporated into a wide range of resource evaluations and risk assessments, informing decisions on how to sustain population dynamics and genetic diversity in plant and animal populations \cite{kareiva1995connecting}, how to most effectively prevent infectious disease spread or reduce risks from wildfire \cite{gray2016applying}, and choices of where to invest in land protection or restoration to help support species range shifts under a changing climate \cite{heller2009biodiversity, littlefield2017connecting, keeley2017habitat}.  

A common requirement for modeling connectivity is a gridded depiction of the landscape in which values for each cell represent some relative value of “resistance” to movement. These resistance grids are developed through several different methods, often involving iterative processes for categorizing resistance weights for different types of barriers based on expert opinion and information on species’ life histories and movement behaviors \cite{spear2010use, zeller2012estimating}. This grid can then be abstracted as a graph \cite{urban2001landscape}, providing a way to quantify ecological distance measures via graph-theoretic metrics. 

The range of mathematical approaches and software tools used for modeling connectivity reflect differences in theoretical approaches, and in the underlying assumptions about how movement proceeds. The classical isolation by distance model (IBD) posits that the least cost distance across the landscape graph acts as a good proxy for ecological distance \cite{wright1943isolation}.  Tools based on this approach typically identify a single “best” route between focal locations, an important result, but one that is highly dependent upon the choice of start and end points, and has limited application if one’s goal is to compare potential options for restoration or protection across a landscape with multiple habitat patches and pathways.  As reviewed in \cite{dickson2019circuit}, from 2006-08, three seminal papers by the late Brad McRae built upon earlier work by \cite{doyle1984random} to demonstrate that isolation by resistance (IBR)\cite{mcrae2006isolation}, which operationalizes the potential for genes and individuals to follow “random walks” across multiple pathways in the same way that electrical current can spread across multiple resistors, can provide an effective tool for considering connectivity potential across landscapes. 

These insights and McRae’s interest in informing conservation applications inspired the Circuitscape software package \cite{mcrae2008using}, which calculates effective resistance and “current flow,” a measure of net movement probability, across heterogeneous landscapes \cite{dickson2019circuit}.  This approach has evolved into a very flexible set of tools that allow users to vary the resistance grid and identification of what can be connected, enabling modelers to address a wide range of questions related to structural (ability of a landscape to support movement) and functional (modeling that is tailored to species-specific traits) connectivity. Through over the past decade Circuitscape has emerged as the most cited landscape connectivity tool in the world \cite{dickson2019circuit}. In the last year alone, Circuitscape has been cited 129 times, and in the past three years, 480 times. Dickson et al. (2019) reviews 277 applications of the software, in fields ranging from gene flow and animal movement to fire, water, and disease spread, and also describes how outputs are being used to inform conservation decisions.

\begin{figure}
    \centering
    \includegraphics[width=\textwidth]{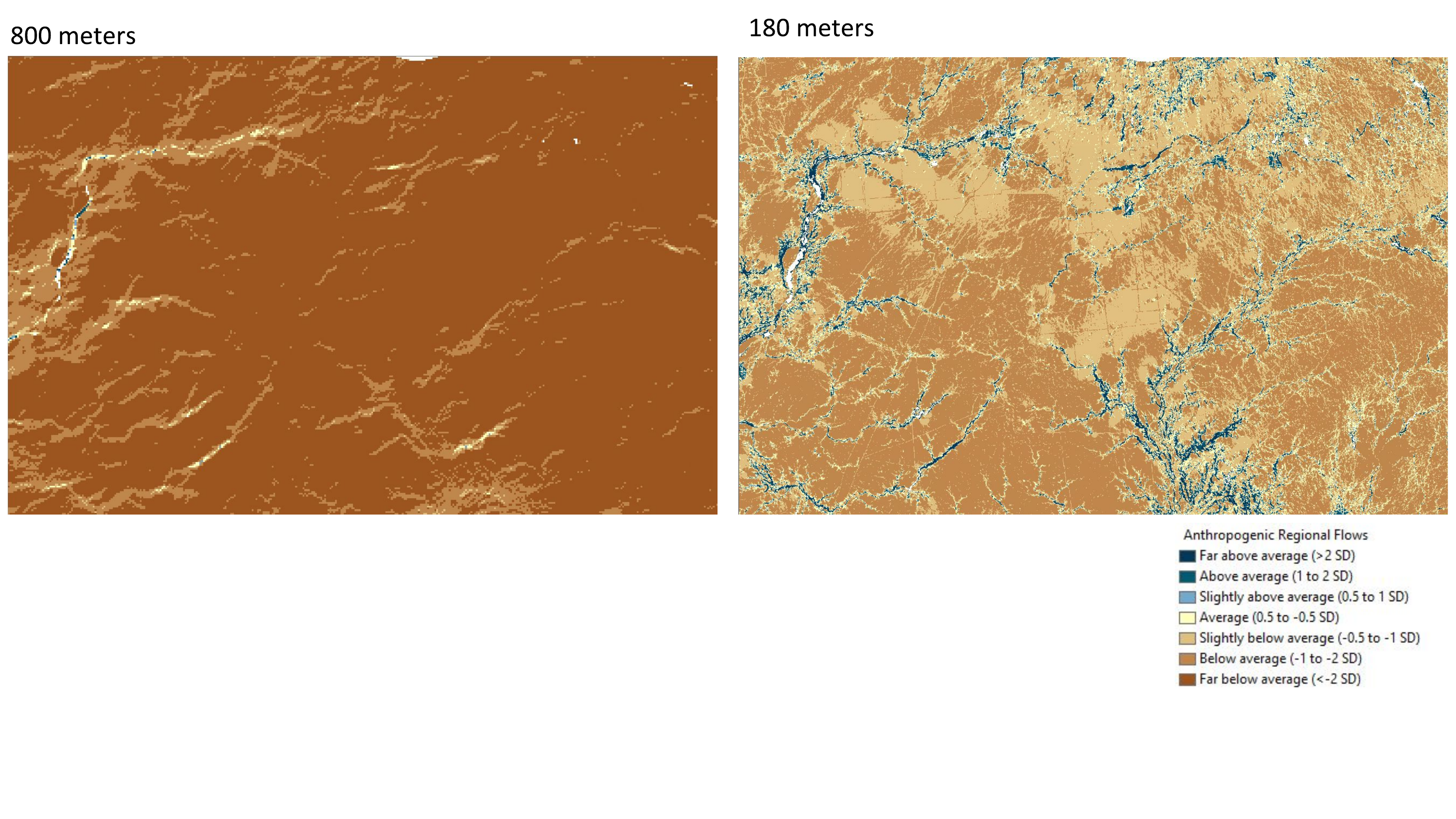}
    \caption{Modellers now require Circuitscape to process increasingly finer resolutions.}
    \label{fig:my_label}
\end{figure}

\section{Reimplementing Circuitscape in Julia}

The goal of increasing the computational power of Circuitscape has been addressed multiple times by developers Brad McRae and Viral Shah.  McRae’s first version was written in Java, before being ported to MATLAB, to improve ease of development. Then, in collaboration with Tanmay Mohapatra, it was translated to Python for flexible scripting, platform independence and release under an open source license. This package makes heavy use of the numpy \cite{van2011numpy}, scipy \cite{jones2014scipy} and pyamg \cite{bell2015pyamg} numerical libraries and solves problems of up to tens of millions of nodes, but connectivity analysis increasingly finer resolution databases often scaled to hundreds of millions of nodes and dozens of focal node pairs, a workload size well beyond its capacity. Circuitscape needed to be upgraded to support the newer demands of user the community. 

To address these demands, McRae and Shah collaborated on  a project called GFLOW \cite{leonard2017gflow}, a software tool for conducting circuit-theory based analyses on supercomputers. Written in the C programmming language and making use of solver libraries such as PetSc \cite{balay2004petsc} and BoomerAMG \cite{yang2002boomeramg}, GFLOW yields state of the art performance on large clusters and supercomputers. However, this performance comes at the cost of accessibility (does not work on the Windows operating system, which the majority of Circuitscape users use), composability (ability to integrate with other software libraries) as well as a tangible difficulty in shipping binaries because of their complicated build processes. For many practitioners and researchers, these constraints are a key barrier, as they prevent GFLOW from being supported on the Windows operating system, and limit integration into a decision support tool or workflow with other tools. 

As the maintainers of the Circuitscape, we saw the need for an open-source implementation that is easy to maintain, is high performance, is accessible to our user base and works well on every platform. We decided to use the Julia programming language \cite{bezanson2017julia}. Julia is an open source dynamic programming language which combines the readability of scripting languages such as R or Python, with the performance of a statically compiled language such as C or Fortran. With relatively little development effort via a straight reimplementation of the algorithm, Julia allowed us to not only significantly improve the package's computational capacity, but provide new user-facing features for free. For example, Julia's first class sparse matrix library and factorization support allowed us to support multiple solvers. Its modern Just in Time (JIT) compiler enables programmers to write generic code by generating specialized code for desired precision, index types and platform. This makes maintenance of the code base simple. Our upgrade, "Circuitscape.jl", is a registered Julia package and is already being used by the community to solve the next generation of connectivity problems.  

By rewriting Circuitscape from scratch in the Julia scientific programming language, our goal is to provide practitioners the ability to model processes like gene flow and species movements at a continental or even global level at relatively fine resolutions, while still being a tool that is open source, widely accessible and easy to use.  We expect that by improving the speed and ability of Circuitscape to handle very large datasets, innovations in areas such as use of new, high resolution datasets, sensitivity testing, and dynamic simulations will soon follow.

The rest of this paper is organized as follows: we present an overview of the algorithm in Section 3, before describing the numerical methods and software packages in Section 4. Section 5 then briefly describe the new features introduced in the upgrade and then present benchmarks both on synthetic problems as well as real user data. We then round off this work with a discussion on how speed improvements enable increased collaboration between computer scientists and ecologists enable improved collaboration between ecologists and conservation managers and other stakeholders. 

\section{Overview of the Algorithm}

\begin{figure}
    \centering
    \includegraphics[width=\textwidth]{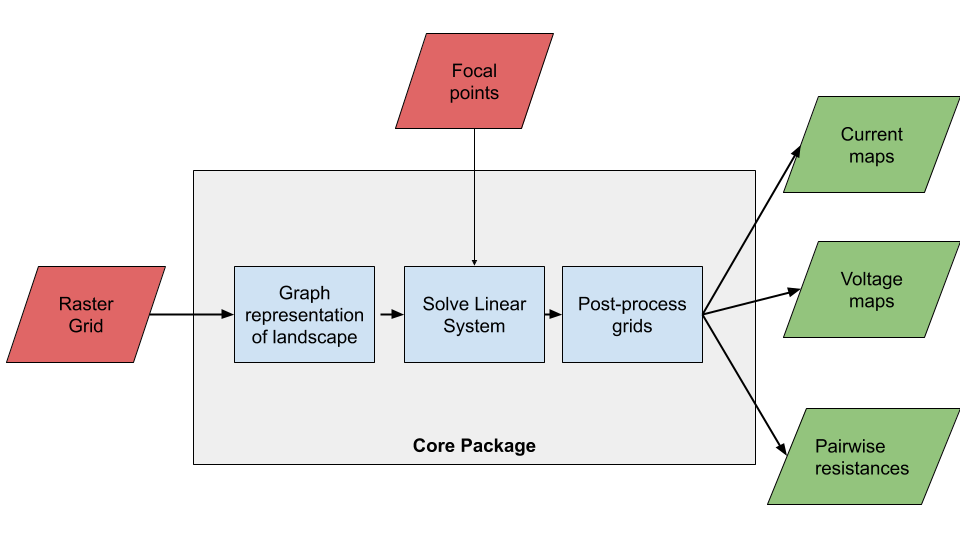}
    \caption{Stages of computation and outputs. The red boxes represent inputs and the green boxes represent output.}
    \label{fig:flow}
\end{figure}

Circuitscape takes as input a raster grid, which in most applications is a spatial discretization of a heterogeneous landscape. Each cell is assigned a value, which represents varying qualities of habitat, dispersal routes, or movement barriers. These “resistance” values are often problem specific, and are either empirically derived based on movement or genetic data, or derived via expert opinion. The following content has been summarized in Figure \ref{fig:flow}. 

\begin{figure}
\centering
\begin{algorithm}[H]
\SetAlgoLined
\LinesNumbered
\KwResult{Current (probability) maps $C$, nodal voltages $V$, pairwise resistances}
 Read input raster grid \;
 Read focal nodes \;
Construct undirected, weighted graph connecting neighboring cells\;
Compute graph Laplacian $G$ \;
\For{all pairs of focal nodes}{
    Set up $I$\;
    Solve linear system $GV = I$ \;
    Compute effective resistance $r$\;
    Write nodal voltages $V$ to disk\;
    Compute nodal currents from $I$ and write current map C to disk\
}

 \caption{Circuitscape - Pairwise Mode }
\end{algorithm}
\end{figure}

This grid can then be represented as a graph, with nodes representing grid cells and edge weights proportional to the movement probabilities or number of migrants exchanged. Other inputs include a list of focal points, which represent points on the landscape of interest to the practitioner. These include habitat patches, protected areas, or populations to connect. These can be specified either through a raster grid with numerical labels for each focal point, or a text file with a list of coordinates. 

Once the graph is constructed, we then compute its laplacian, which is an alternate representation of the landscape. We then solve for Ohm’s law using the computed graph laplacian as the conductance matrix and the node locations of the focal points as current sources and sinks. On solving the system, we obtain a list of nodal voltages, which are then used to compute branch currents, which represent movement probabilities along branches of the graph. The ecological significance of these various quantities are summarized in related works \cite{mcrae2008using}. The branch currents are then accumulated on each node as nodal currents, which are then written to disk as a raster ASCII grid, for easy import into a geographic information system (GIS). 

\section{Overview of Numerical Methods}

The vast majority of the computation in Circuitscape is applying Ohm’s law and Kirchoff’s law over very large resistive grids, and solving a large sparse linear system: 
$$
GV = I
$$
where $G$ is the graph laplacian representation of the landscape is stored as a large sparse matrix, $I$ are the current sources, and $V$ are the nodal voltages. We provide two solver options to the user: one based on the choleksy factorization and a preconditioned iterative method.

The Cholesky factorization \cite{higham2009cholesky} is efficient for applications with smaller study areas and a large number of focal nodes. The sparse matrix is factored once and the solution to multiple pairs is computed via back substitution. Since the cost for backsubstitution is polynomially smaller than the cost of factorization, problems with a large number of focal pairs scale efficiently.

The solver that scales to large problems is a preconditioned conjugate gradient (PCG) \cite{trefethen1997numerical}, with an algebraic multigrid preconditioner (AMG) \cite{vanvek1996algebraic}. We leverage Julia implementations of these methods from open source packages IterativeSolvers.jl and AlgebraicMultigrid.jl.

\subsection{Preconditioner Implementation}

Multigrid preconditioners are often used in problems solved across large spatial domains. In general, they take the original grid and generate a hierarchy of coarser grids, via predefined or algebraically derived restriction and interpolation operators. Large problems are solved by restricting quantities down to the coarsest level, obtaining a fast solution, and then interpolating the solution back to the original size. This procedure is often referred to as a “V” cycle.

The efficacy of a good multigrid preconditioner is to with obtaining good restriction and interpolation operators. Usually, these operators are derived from the structure of the underlying problem. However, one can also estimate these operators algebraically, from the numerical values at each grid cell. This procedure is often referred to as Algebraic Multigrid (AMG). There are several variants of AMG, and they are summarized in \cite{stuben2001review}. We use a variant called the Smoothed Aggregation AMG, which is known to work well for solving matrices generated by elliptic partial differential equations, such as Laplace’s equation. These laplacian matrices are structurally identical to the laplacians of planar graphs that are generated in Circuitscape.

\section{New Features}

\subsection{New Solver Mode}

Circuitscape now supports a new solver based on the Cholesky factorization of the graph laplacian representation of the landscape. We use the SuperNodal sparse cholesky factorization implemented in the CHOLMOD library \cite{chen2008algorithm}. The Julia programming language has first class sparse matrix library and matrix factorization support, which makes it easy to use this solver with a single function call.

This solver mode also enables new approaches that divide the landscape into tiles and process them independently \cite{pelletier2014applying}or analyses that move sliding windows across the landscape \cite{mcrae2016conserving}, both of which yield fine scale and localized results. 

The obvious limitation of this solver, as is of most sparse direct solvers, is that they work very well for relatively small matrices and run out of memory for larger ones. This exponential growth in memory is a by product of the factorization itself, as factors are often less sparse than the original matrix. 

\subsection{Generic Programming}

Julia supports the development of generic code while relying on the compiler to generate specialized code for different platforms and precision required. Generic programs are powerful tools for our users, and can be made to adapt to their requirements. For example, the python package had a 32-bit integer type for indexing matrices hardcoded throughout the package, which limited the size of the computation the package ran. This led to crashes when running simulations for hundreds of millions of nodes. One such dataset studying the Mojave desert tortoise \cite{gray2019} used a resistance surface of 437 million pixels. This crashed the old version of Circuitscape, while the upgrade ran smoothly to completion. 
 
Circuitscape.jl defaults to a 32-bit integer type for cache efficiency, but provide users with a runtime flag to switch to 64-bit integer indexing. Our Julia code was written generically, and this feature came for free. This allowed us to scale to problem sizes of hundreds of millions. Upgrading the python package to support this feature would require significant effort. 

\subsection{Improved Parallel Computations}


Parallel computing in Circuitscape.jl is inherently faster, because the preconditioner can be serialized as a byte stream and sent over the network to other processes. This is not possible in the python code, as the python multiprocessing module internally uses the package pickle to serialize objects, and pyamg objects, implemented in C++, are not “pickle-able” \cite{pydoc}. The ability to serialize native Julia objects significantly reduced the amount of effort to parallelize Circuitscape.

In addition to faster parallel processing, we also extend parallelism support for more problem types in Circuitscape. In the benchmarks section, a dataset which received speed improvements from these new features. 

The Julia version also lets the user call Circuitscape itself in parallel. As users aim to run Circuitscape over entire countries or continents, landscape are often divided into multiple tiles \cite{pelletier2014applying}. This feature allows users to process (a batch of ) these tiles by running different Circuitscape problem instances on each tile in parallel. 

These parallel features work on all platforms: Linux, MacOS and Windows. 

\section{Benchmarks}

\subsection{Experimental Setup}
We conducted our experiments on an Intel(R) Xeon(R) Silver 4114 CPU @ 2.20GHz with 384 GB of RAM. We used Julia v1.1.0 and Circuitscape v5.5.0, and compared against Circuitscape v4.0.5 

\subsection{Standard Benchmarks}
We benchmarked Circuitscape on standard synethic problems. These datasets can be found at \url{https://github.com/Circuitscape/BigTests/}. 

\begin{figure}
    \centering
    \includegraphics{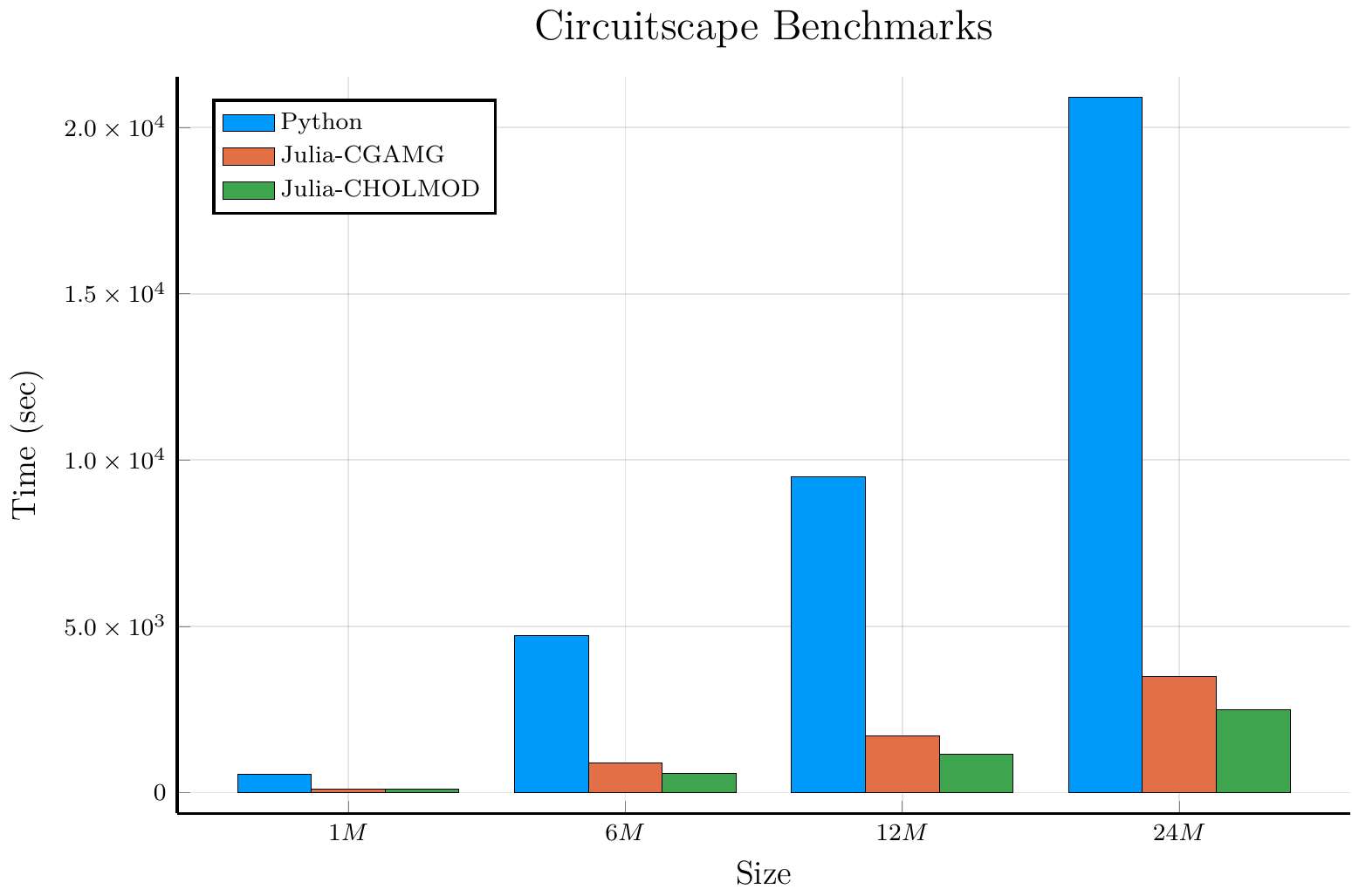}
    \caption{Results on the standard Circuitscape benchmark suite. The red column standards for the Julia package run under the AMG solver and green column stands for the Julia package run with the Cholesky-based CHOLMOD solver.}
    \label{fig:my_label}
\end{figure}

\subsection{Benchmarks on Real Data}

We ran the Python and Julia version of Circuitscape on a dataset used to model connectivity in the Sonoran desert \cite{drake2017using}. We used to all-to-one scenario \cite{mcrae2009circuitscape}, which models species migrating across the landscape from many different areas to one area. When the researcher used the old version of Circuitscape, this mode did not support parallelism, and the entire computation took over 2 days to run. Parallelism was trivial to support in the Julia version, which reduced execution time to less than 3 hours, resulting in a speedup of nearly 18x.



\section{Discussion}

The growth in popularity of circuit theory to understand ecological processes has to lead to widespread adoption of the Circuitscape software package. \cite{mcrae_shah_mohapatra_anantharaman}. The Circuitscape project has always evolved with the demands of the users and this upgrade to the Julia programming language seeks to drive the next generation of compute-intensive connectivity models.
Users no longer have to coarsen their analysis due to limitations of their tooling.


This advance also enables users to experiment with different scenarios and perform sensitivity analysis, which is used by most users to understand and validate the effect of all the individual layers that go into a resistance map. With previous versions, practical sensitivity testing using Circuitscape was prohibitively slow. This upgrade helps users rapidly prototype their resistance maps.

Cutting the execution time from weeks to days or from days to hours also enables stronger stakeholder engagement, improving the feedback loop with policy makers, resulting in faster turnaround time for conservation decisions. 

The Julia programming language's ease of use and mathematical syntax enables fast scripting, while its modern compiler allows quick and dirty prototypes to scale to large datasets. Julia is already being used by ecologists to model ecological networks \cite{timothee_poisot_2018_1438428} and bioenergetic food webs \cite{delmas_eva_2019_2584373}, and will continue to enable advances in other areas of computational ecology. 


\section{Conclusion}
We present an upgrade to the Circuitscape package, which will allow researchers to analyze ecological processes over large landscapes at fine resolutions. Our upgrade in the high performance Julia programming language presents upto a 1800\% improvement in computation time and the ability to solve landscapes with billions of pixels. Julia’s sophisticated compiler allows for faster parallelism, generic programming and composability with other software packages.

Circuitscape.jl is open source and is available at \url{https://github.com/Circuitscape/Circuitscape.jl} under the MIT license. Binaries are available on \url{https://circuitscape.org/downloads/}. It can also be installed using the Julia package manager by starting Julia, and entering \texttt{using Pkg; Pkg.add("Circuitscape")}. 

\section{Acknowledgement}
The work to update Circuitscape to the Julia computing language was funded through a grant from NASA’s Ecological Forecasting Program (16-ECO4CAST0018), and a grant from the Wilburforce Foundation.  The authors also wish to acknowledge the Wilburforce Foundation’s long-term investment in Circuitscape and related tools through several previous awards to Dr. Brad McRae.  We thank all of the Circuitsape users who have shared datasets, reported issues, and provided feedback on various versions of these tools, and look forward to continued collaboration with the user community.

\bibliography{references}  

\end{document}